\documentclass[aip,amsmath,amssymb,reprint,]{revtex4-1}

\usepackage{graphicx}
\usepackage{dcolumn}
\usepackage{bm}
\usepackage[utf8]{inputenc}
\usepackage[T1]{fontenc}
\usepackage{mathptmx}
\usepackage{etoolbox}

\makeatletter
\def\@email#1#2{
 \endgroup
 \patchcmd{\titleblock@produce}
  {\frontmatter@RRAPformat}
  {\frontmatter@RRAPformat{\produce@RRAP{*#1\href{mailto:#2}{#2}}}\frontmatter@RRAPformat}
  {}{}
}
\makeatother

\begin{document}

\preprint{AIP/123-QED}

\title[Sample title]{Research on quantum compilation of neutral atom quantum computing platform}

\author{Xu Chongyuan$^\dagger$}
\email{moke2001@whu.edu.cn}
\affiliation{ 
Wuhan University, College of Physical Science and Technology, Wuhan, 430072, China
}

\date{\today}

\begin{abstract}
Quantum compilation is the process of decomposing high-level quantum algorithms or arbitrary unitary operations into quantum circuits composed of a specific set of quantum gates. Neutral atom quantum computing platform is a quantum computing implementation method with high controllability and scalability, but its quantum compilation method is not mature. We systematically review the quantum compilation methods based on matrix decomposition, and propose a compilation algorithm suitable for neutral atom quantum computing, which can effectively decompose any unitary operation into a series of quantum gates suitable for the neutral atom platform, and ensure that the generated quantum circuits can run directly on the platform.
\end{abstract}

\maketitle


\section{Introduction}

Quantum computing is a novel scheme of information processing and computation based on the principles of quantum mechanics, using the states and interactions of quantum systems. Quantum computers have broad application prospects, such as: quantum simulation, which uses quantum computers to simulate complex quantum systems, such as molecules, materials, chemical reactions, etc., to reveal their microscopic structure and properties, and provide new tools and methods for scientific research and engineering design \cite{aspuru2012photonic}\cite{georgescu2014quantum}; quantum algorithms, which use quantum computers to execute some algorithms that surpass the classical computational capabilities, such as Shor’s algorithm \cite{shor1994algorithms}, Grover’s algorithm \cite{grover1996fast}, etc., to achieve factorization of large numbers, database search, number theory research, etc.
At the physical level, there are different ways to implement a quantum computation. Some common ones are: superconducting quantum computer, which uses the Josephson structure in the superconducting circuit to realize the quantum coherent oscillation between two energy levels, as the carrier of quantum bits \cite{clarke2008superconducting}\cite{devoret2013superconducting}; ion trap quantum computer, which uses laser field to manipulate ionized atomic nuclei or electrons, to realize the quantum coherent control of their internal energy levels or external motion states, as the carrier of quantum bits \cite{leibfried2003quantum}\cite{monroe2013scaling}. In recent years, the neutral atom system with controllable interaction, long coherence time and strong scalability is one of the powerful candidates for realizing quantum computer, and the research on quantum computation based on neutral atom system is flourishing \cite{brubaker2022versatile}.
At the abstract level, in order to fully utilize the computational power of quantum systems, a universal quantum computer is necessary. A universal quantum computer is a device that can implement any unitary operation or approximate any unitary operation to any precision. However, in neutral atom platform or other specific physical systems, only a few specific unitary operations can be used, which is similar to the basic logic gates in classical computer. Therefore, we need to decompose the applied unitary operation into a sequence of experimentally available unitary operations (i.e., universal quantum gates), which is called quantum compilation. If we can find such a decomposition for any unitary operation acting on n quantum bits, then such a set of quantum gates is called a universal quantum gate set for n quantum bits.
Since different physical systems implement different universal quantum gate sets, it is necessary to design unique methods for different physical systems, so that the quantum compilation program can output a sequence of quantum gates that can be executed on the quantum computing hardware. Any quantum computing system has a limitation on the quantum volume, which is the size of the largest square quantum circuit that they can successfully implement. As of February 2023, the highest record of quantum volume is 215, achieved by Quantinuum’s ion trap quantum computer \cite{moses2023race}. Therefore, for a specific unitary operation, we hope that the quantum circuit generated by quantum compilation is as small as possible.
This paper mainly studies the quantum compilation algorithm based on matrix decomposition and proposes an adaptation method for neutral atom quantum computing platform. The rest of this paper is organized as follows:
In Section 2, we introduce the basic principles of neutral atom quantum computing platform and quantum compilation, which are the basis of our work; in Section 3, we introduce the matrix decomposition-based quantum compilation algorithm and our adaptation method for neutral atom quantum computing platform; in Section 4, we apply our quantum compilation algorithm to neutral atom quantum computing platform and analyze its performance; finally we summarize and outlook our work. Our adaptation software has been posted on Github.


\section{Basic Principles}

\subsection{Quantum Computing and Quantum Compilation}

Quantum computation on neutral atom system is one of the fast-developing physical implementation schemes of quantum computation in recent years. In this system, the quantum computation states $|0\rangle$ and $|1\rangle$ are encoded in two magnetic sublevels of the ground state of alkali metal (usually rubidium or cesium) \cite{Graham2022Multi}. Implementing a quantum computation on neutral atom platform can be summarized into five steps: cooling of neutral atoms, trapping of neutral atoms, initialization of quantum bit states, implementing a series of single quantum bit gates or entanglement quantum gates, and finally measurement.

The first step is to cool the alkali metal gas from the atomic oven. Neutral atom quantum computation requires cooling the atomic cloud to sub-microkelvin temperature, to reduce the atomic motion and collision, thereby improving the quantum coherence. The cooling process is generally divided into two stages: laser cooling and magneto-optical trap cooling. Laser cooling is to use the Doppler effect and radiation pressure between laser and atom, to use red-detuned light to make atoms tend to absorb and emit photons from the opposite direction of their own motion, thereby reducing the atomic kinetic energy \cite{metcalf1999laser}. Magneto-optical trap cooling is to use the combined effect of magnetic field and laser, to form a spatial potential trap, to confine atoms in the trap, and further cool them by methods such as adiabatic demagnetization \cite{chu1998nobel}. These two stages can cool the atomic cloud to the order of millikelvin temperature.

After the above process, we can obtain a large number of low-temperature alkali metal atoms in the center of a high-quality vacuum cavity. Then we need to fix these atoms at certain positions by optical lattice or optical tweezer, to form a neutral atom array \cite{endres2016atom}. As shown in Figure 1(a), optical tweezer is to use a focused laser beam to form a tiny potential trap in space, to capture atoms in the trap, and control the atomic position by adjusting the intensity and frequency of the laser. Optical tweezer can realize precise operation and manipulation of single or multiple atoms, thereby constructing the platform of neutral atom quantum computation. Constructing an array using optical tweezer generally consists of two steps: loading and arranging. The cooled atomic cloud is injected into a two-dimensional or three-dimensional optical tweezer array. Under the effect of collisional blockade \cite{Wang2023Controllable}, there is at most one atom in each optical tweezer; by moving or merging optical tweezers, we can arrange neutral atoms according to the desired geometric shape. These two steps can realize addressing operation of neutral atom quantum bits.

Initialization is the process of putting neutral atom quantum bits into a known initial quantum state for subsequent quantum operations and measurements. The initialization process generally uses laser pumping technology \cite{cohen1998nobel}, that is, using polarized laser to selectively excite or de-excite atoms, making atoms eventually reach a certain energy level state. Due to the effect of selection rules, after a period of laser action, atoms will all populate on the dark state. Since the energy level structure of the same kind of atoms is consistent, using the same pumping light can simultaneously realize state initialization of all atoms in the array.

Implementing single quantum bit gate is to perform single quantum bit operation on neutral atom quantum bit, usually using microwave or Raman laser. As shown in Figure 1(b), microwave is to use electromagnetic wave corresponding to energy level difference to drive quantum bit \cite{Kuhr2005Analysis}, thereby realizing state transition between quantum bits; Raman transition \cite{Yavuz2006Fast} can control population and phase of quantum state by controlling Raman laser action time and phase. The single bit gate on neutral atom system is a $C(\theta,\phi)$ gate, where $\theta$ is controlled by laser action time and $\phi$ is controlled by laser phase. The form of $C(\theta,\phi)$ gate is as follows:

\begin{equation}
C(\theta,\phi)=
\begin{pmatrix}
\cos\frac{\theta}{2} & -e^{i\phi}\sin\frac{\theta}{2}\\
e^{-i\phi}\sin\frac{\theta}{2} & \cos\frac{\theta}{2}
\end{pmatrix}
\end{equation}

Implementing multi-quantum bit gate is to perform unitary operation on multi-quantum bit that generates entanglement between quantum bits, usually using interaction between atoms. There are two main mechanisms of interaction between atoms: collisional interaction \cite{jaksch1999entanglement} and Rydberg interaction. Currently, entanglement is widely realized by using Rydberg dipole-dipole interaction \cite{walker2012entanglement}, as shown in Figure 1(c), this interaction mode utilizes the dipole moment interaction between atoms. In the experiment, atoms are excited by Rydberg laser. Under the influence of Rydberg blockade \cite{Jaksch2000Fast}, the two-atom dark state formed by the resonant dipole-dipole interaction of atoms undergoes adiabatic evolution \cite{petrosyan2017high}, thereby changing the energy level and phase of the two atoms. Dipole interaction can realize controlled phase gate ($CZ$ gate or $CCZ$ gate) between two quantum bits or three quantum bits.

\begin{figure}[!htbp]
    \centering
    \includegraphics[width=0.5\textwidth]{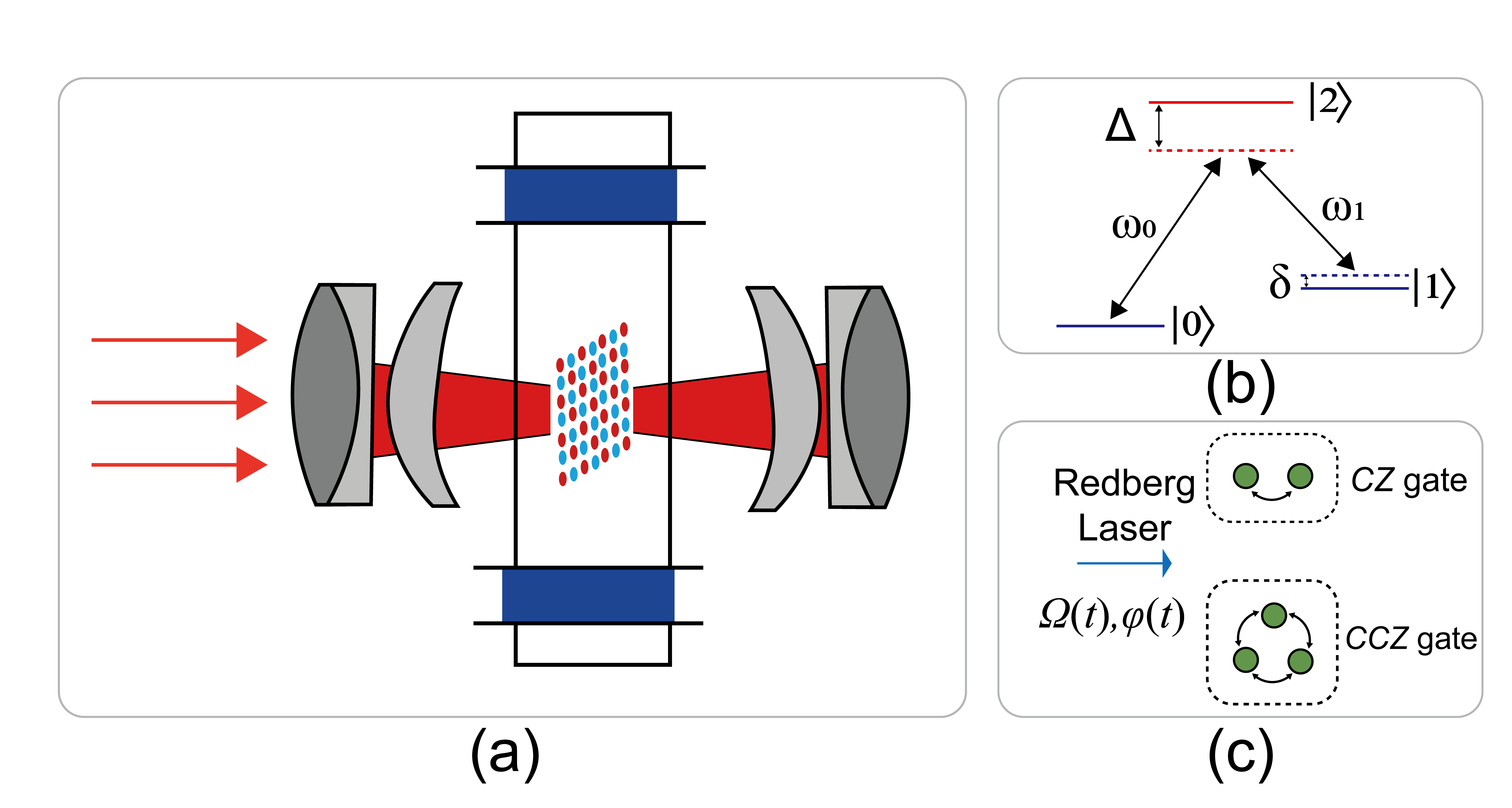}
    \caption{The implementation process of neutral atom quantum computation system. Figure (a): The overall experimental structure, the neutral atom array is placed in a high-quality vacuum cavity, and the microwave or laser used for initialization and manipulation of quantum bits are incident into the vacuum cavity through a high numerical aperture lens; Figure (b): The process of implementing single quantum bit gate by Raman transition, due to the restriction of selection rules, Raman process is a two-photon process, thereby making transitions between two energy levels that do not satisfy in the ground state; Figure (c): The process of implementing entanglement quantum gate by Rydberg dipole-dipole interaction and Rydberg blockade, under the action of Rydberg laser, we can realize two-atom $CZ$ quantum gate or three-atom $CCZ$ quantum gate between atoms.}
\end{figure}

Currently, the fidelity of single quantum bit gate can reach a very high level, about $0.9962(16)$, and the crosstalk between different single quantum bit gates is less than $0.002$\cite{wang2016single}; the fidelity of entanglement gate on neutral atom quantum computation system can reach $99.5\%$\cite{Evered2023High}, and the fidelity of entanglement gate on "Hanyuan" series neutral atom quantum computer of Chinese Academy of Sciences Institute of Precision Measurement is about $98\%$.

Finally we need to perform measurement on the system, measurement is the process of obtaining state information of neutral atom quantum bit, usually using fluorescence detection technology \cite{bakr2009quantum}, that is, using resonant laser to heat atoms in $|1\rangle$ state, making them lose from optical tweezer. Repeatedly use high numerical aperture lens to collect fluorescence generated by atomic spontaneous emission, and detect fluorescence signal by devices such as photomultiplier tube or charge coupled device. The improved strategy is to use high numerical aperture lens to improve atomic fluorescence collection efficiency, and carefully optimize detection light frequency and resonance with closed transition energy level to suppress Raman transition, this way can achieve more than $97\%$ detection efficiency and atomic loss rate less than $2\%$\cite{kwon2017parallel}.

\subsection{Quantum compilation}

Quantum computation consists of the following three processes:

First, initialize the quantum bits: The process of initializing the quantum computation system is to adjust the state of the quantum bits or quantum system to a known initial state, usually $|0\rangle$ or $|+\rangle$, for subsequent quantum operations and measurements \cite{matthews2021get}.

Second, apply a series of operations on the quantum bits: In quantum computation, the way to operate on the quantum bits is to use some basic quantum gates. Quantum gates are unitary matrices that perform linear transformations on quantum state vectors, such as Hadamard gate, Pauli gate, CNOT gate, etc. A quantum circuit is a series of quantum gates arranged in order, which can implement various different unitary operations. We have listed some common quantum gates in Figure 2, where Figure 2(a) shows three general rotation gates; Figure 2(b) shows three parameter-free single-qubit gates; Figure 2(c) shows the $CNOT$ gate that acts on two qubits; and Figure 2(d) shows the most general form of a single-qubit gate on the Bloch sphere.

Finally, measure the quantum system: In quantum computation, to obtain information of a quantum bit or a quantum system, we need to measure it. Measurement is an irreversible process, which will cause the quantum state vector to collapse to a certain basic state and output the corresponding result. This process follows the generalized statistical interpretation.

The part of interest in quantum compilation mainly focuses on the second step of the above process, that is, how to decompose a unitary operation into a series of basic quantum gate operations. The universal quantum gate sets used in common quantum compilation algorithms include $R_x$ gate, $R_y$ gate, $R_z$ gate and $CNOT$ gate. The first three are single quantum bit gates that rotate around a fixed axis on the Bloch sphere. The $CNOT$ gate is an entanglement quantum gate that acts on two quantum bits. In addition, in discrete quantum compilation, Clifford+T gate set is also widely used \cite{bravyi2016improved}. The entanglement operation in this gate set is also $CNOT$ gate. The other single quantum bit gates include $T$ gate, $S$ gate and $H$ gate which is Hadamard gate.

\begin{figure}[!htbp]
    \centering
    \includegraphics[width=0.5\textwidth]{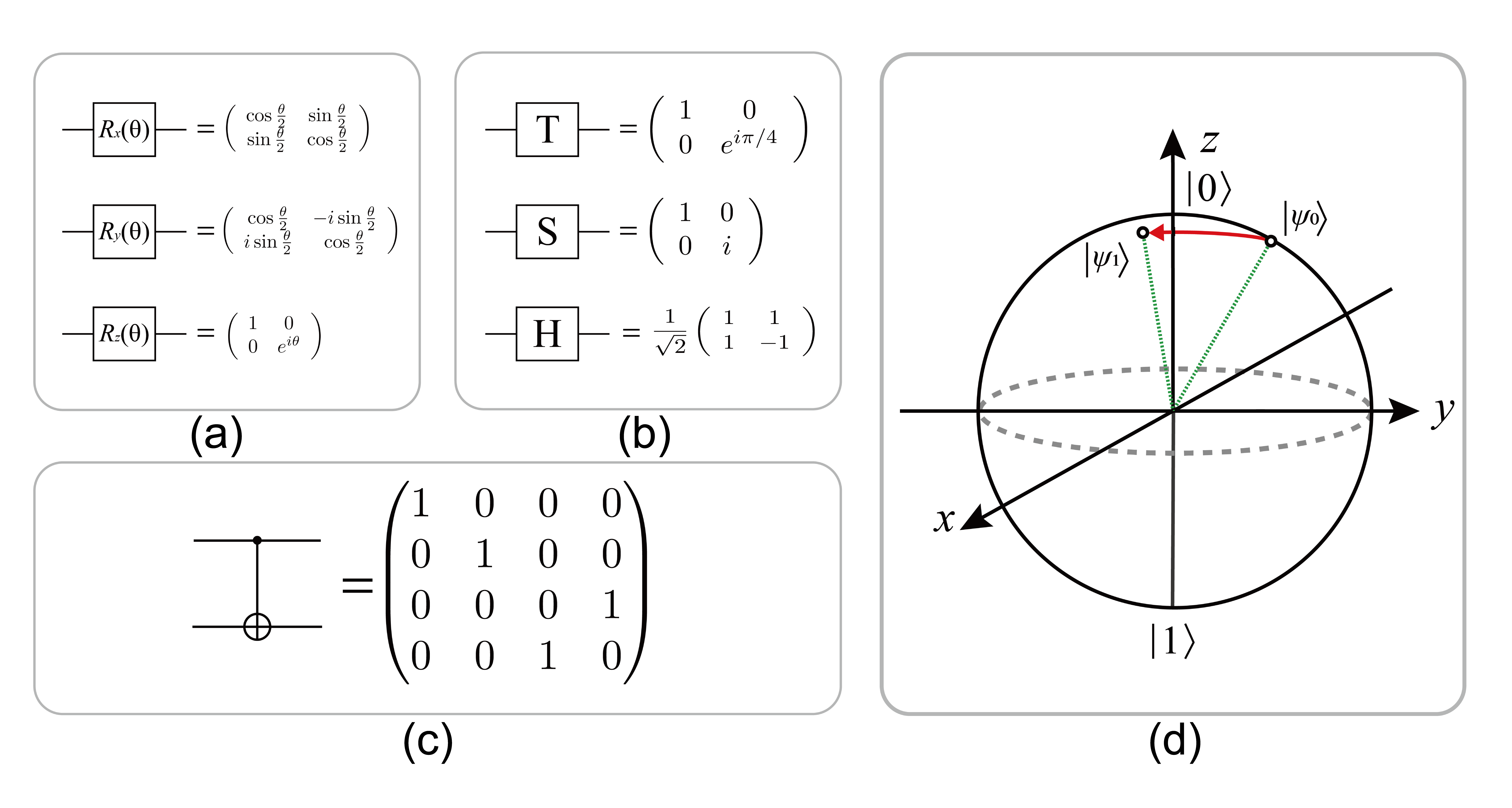}
    \caption{Some concepts related to quantum computation. Figure (a): The circuit representation and matrix form of three parameterized rotation gates; Figure (b): The circuit representation and matrix form of three gates in Clifford+T gate set; Figure (c): The commonly used entanglement gate $CNOT$ gate; Figure (d): The form of Bloch sphere, the state of a single quantum bit is represented by a point on the Bloch sphere, and the operation of a quantum gate corresponds to a rotation around a fixed axis on the Bloch sphere.}
\end{figure}

Common quantum compilation algorithms have two methods: discrete approximation and matrix decomposition. The method of quantum compilation based on matrix decomposition is to use QR decomposition, quantum Shannon decomposition, SVD decomposition, etc., to decompose any unitary matrix into a series of basic matrices, equivalent to a series of basic quantum gates, such as single quantum bit gates and controlled not gates \cite{krol2022efficient}. The advantage of this method is that it can accurately implement any unitary matrix without introducing approximation error; the disadvantage of this method is that the decomposition process may be time-consuming, and the generated quantum circuit may be long, thereby increasing the execution time and error rate. The method of quantum compilation based on discrete approximation is to use some algorithms, such as Solovay-Kitaev algorithm \cite{kitaev2002classical}, Kitaev-Shen-Vyalyi algorithm \cite{hattori2018quantum}, etc., to approximate any unitary matrix as a series of basic quantum gates, such as Clifford gates and T gates. The advantage of this method is that it can quickly generate quantum circuits, and the generated quantum circuits can adapt to different quantum computer architectures. The disadvantage of this method is that the approximation process will introduce some errors, and to achieve higher accuracy requires longer quantum circuits. In addition, with the development of artificial intelligence technology, many quantum compilation methods based on artificial intelligence techniques such as machine learning have been proposed \cite{moro2021quantum}.

The discrete approximation-based quantum compilation algorithm will generate longer quantum circuits. Currently, the main algorithm used for neutral atom quantum computing platform is still based on matrix decomposition-based exact compilation algorithm. Previously some adaptation algorithms for neutral atom quantum computing platform were proposed \cite{martinez2016compiling}. These algorithms had to use optimization algorithms such as quasi-Newton algorithm or simulated annealing algorithm when decomposing entangled quantum operations, and the compilation might fail. We think that the length of the quantum circuit generated by the QSD and other matrix decomposition-based quantum compilation algorithms is close to the lower bound of exact decomposition. The key problem is that the single quantum bit gates in the universal quantum gate set of these algorithms are ${R_x(\theta),R_y(\theta),R_z(\theta)}$, which are different from the single quantum bit gate $C(\theta,\phi)$ on neutral atom quantum computing platform. We have found a method based on quaternions that can efficiently synthesize the single quantum bit gates in the quantum circuit generated by the matrix decomposition-based quantum compilation algorithm into several single quantum bit gates $C(\theta,\phi)$. This method reduces the circuit length and obtains a quantum circuit that can be directly executed on the neutral atom quantum computing platform.


\section{Quantum compilation based on matrix decomposition}

The quantum circuit synthesis method based on matrix decomposition is one of the important methods of quantum compilation, which can map any unitary operation to any quantum gate set. The effective implementation of this decomposition will transform a general unitary operation into a sequence of basic quantum gates, which is the key to execute these algorithms on existing quantum computers. This decomposition can provide optimization for the whole quantum circuit, and also test some algorithms on the quantum computing platform. In this part, we will introduce the quantum compilation algorithms based on QRD algorithm and QSD algorithm. Before that, we need to review some basic principles.

\subsection{Basic principles}

Barenco et al. published an important paper on quantum compilation in 1995 \cite{Barenco1995Elementary}, in which they proved that the gate set composed of all single quantum bit gates and $CNOT$ gates is universal, that is, any multi-bit unitary transformation $U(2^n)$ can be exactly implemented by the combination of these gates. They studied how to use these basic gates to implement other gates, such as Toffoli gate, etc., and the number of basic gates required. These gates play a core role in the construction of many quantum computation networks. The most influential ones for quantum compilation work are the universal implementations of single quantum bit gate, single control bit controlled gate and multi control bit controlled gate. The decomposition of these quantum operations is shown in Figure 3.

\begin{figure}[!htbp]
    \centering
    \includegraphics[width=0.5\textwidth]{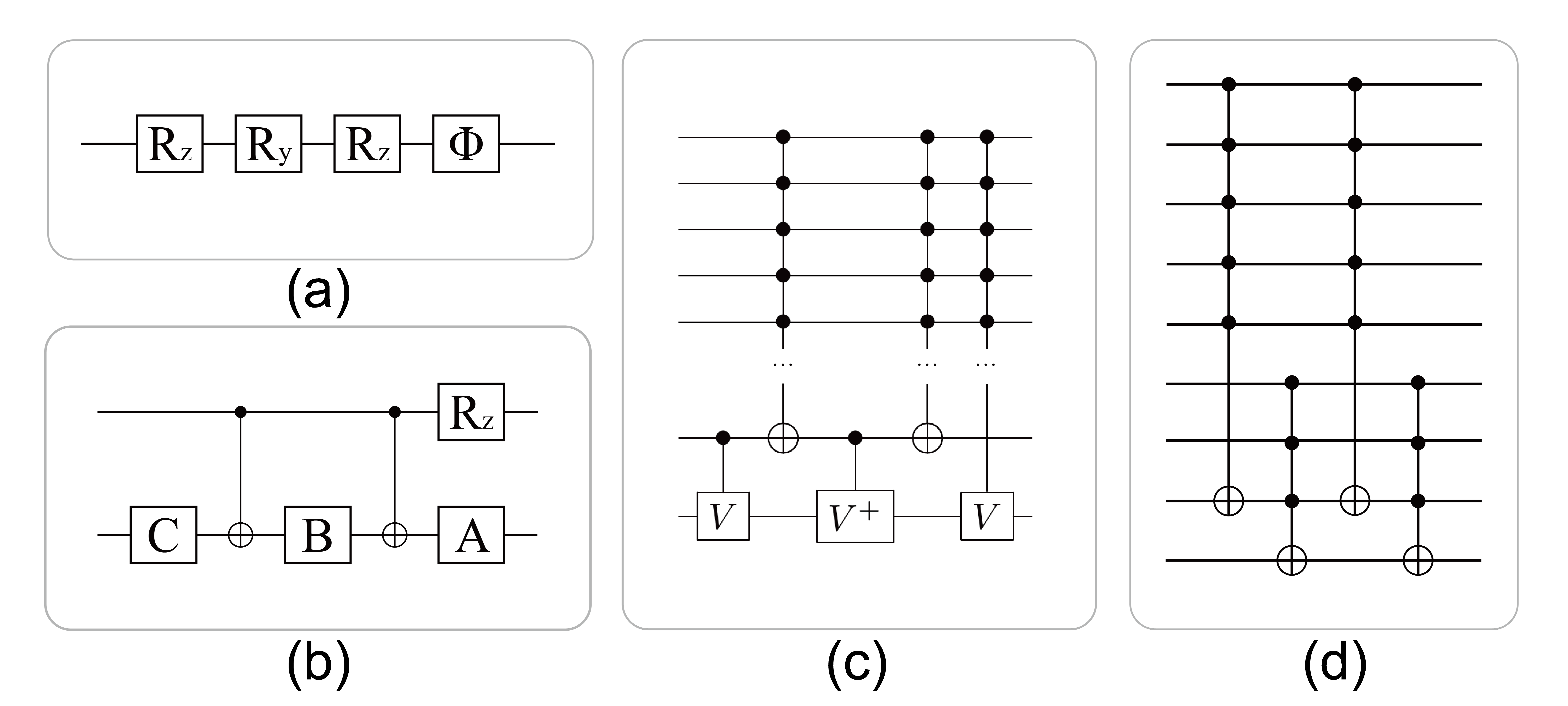}
    \caption{Barenco configuration. Figure (a): The implementation of single quantum bit gate, the corresponding parameters can be obtained according to the form of corresponding single quantum bit gate; Figure (b): The implementation of single control bit controlled gate, where $A,B,C$ three single quantum bit gates satisfy $ABC=I$ and $A\sigma_xB\sigma_xC=U$, $U$ is the target operation; Figure (c): Implementing a multi control bit controlled gate by recursion, the end point of recursion is $CU$ gate or $CNOT$ gate; Figure (d): When the number of quantum bits $n\geq5$ and the number of control bits $m\le n-5$, a ladder-type multi control bit controlled not gate with fewer control bits can be constructed, which has higher decomposition efficiency than recursive decomposition.}
\end{figure}

As a theoretical basis and a practical tool for quantum compilation, this work has had a profound impact on subsequent quantum compilation research. Our work will also follow the methods provided by this literature for the implementation of basic quantum gates such as $C^nU$. In addition, the work on optimal construction on small number of quantum bit circuits \cite{shende2003minimal} will also be applied in our work.

\subsection{QRD algorithm}

QRD algorithm is an efficient algorithm based on matrix decomposition method, which decomposes any n-bit unitary operation into basic quantum gates. QRD algorithm consists of three parts: QR decomposition, GCB encoding and control bit elimination. The process of decomposition by stages is shown in Figure 4.

\begin{figure}[!htbp]
    \centering
    \includegraphics[width=0.5\textwidth]{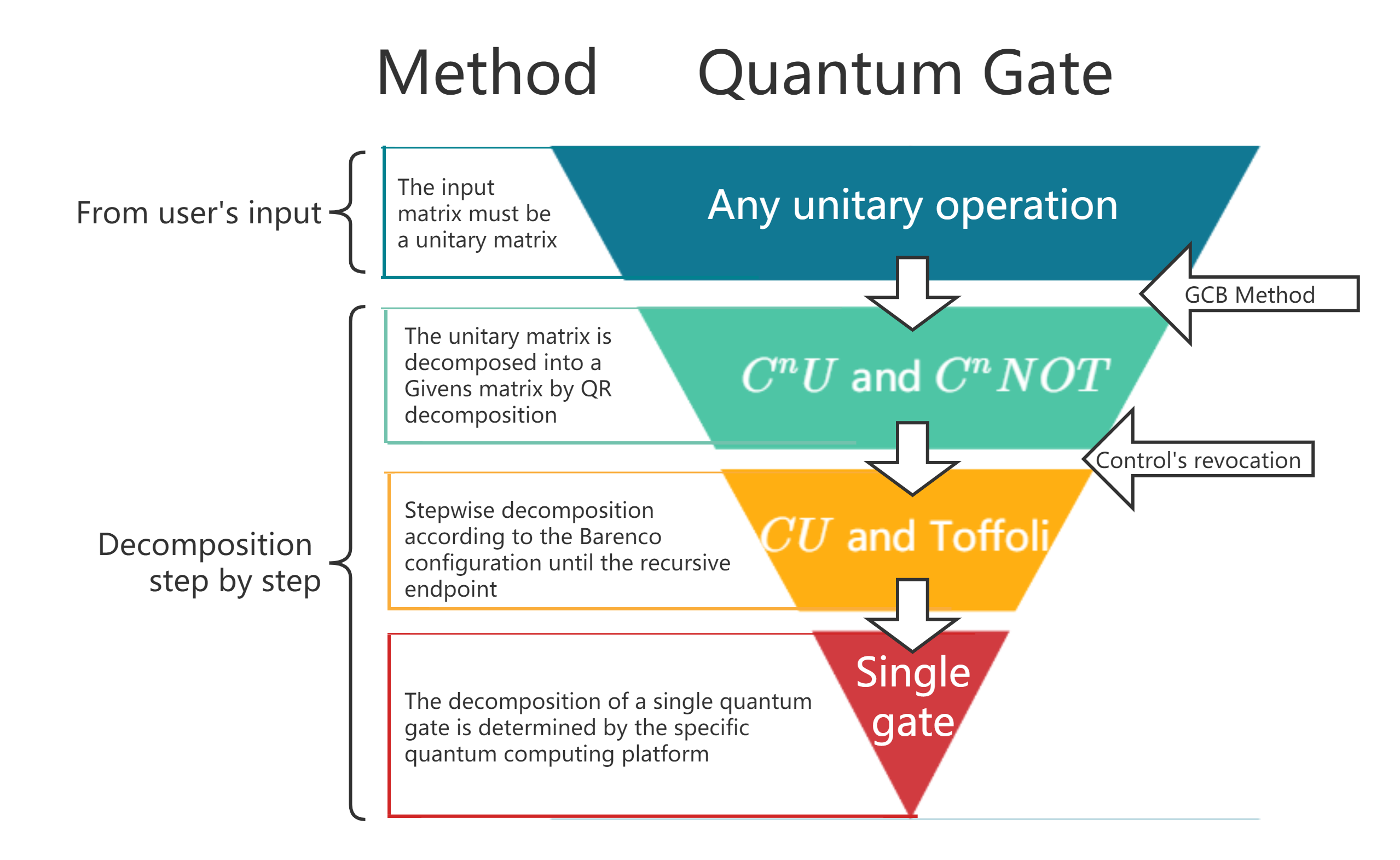}
    \caption{The process of implementing a unitary operation decomposition using QRD algorithm. The whole process is a process of decomposition by stages. QRD algorithm includes two optimization modules: GCB encoding and control bit elimination.}
\end{figure}

QR decomposition is the core step of QRD algorithm \cite{vartiainen2004efficient}, which decomposes n-bit unitary operation into a product of several Givens rotations. Givens rotation is a two-dimensional rotation in unitary space, which can be represented in matrix form. In the decomposition process, we may need to use the extension of quantum gates, which is the process of tensor product of low-dimensional quantum operation and unit operation. The specific operation process is shown in Figure 5(a).

\begin{equation}
\prod^{2^n-1}_{i=1}\prod^{2^n}_{j=i+1}G^{2^n-i}_{j,j-1}U=I
\end{equation}

Where $G^{2^n-i}_{j,j-1}$ represents a Givens matrix, which rotates between the basis vectors $|j-1\rangle$ and $|j\rangle$, so that the element in the i-th column of the j-th row on the right side of the target matrix can be eliminated, and the result is still a unitary matrix after the action. The process of eliminating elements starts from the lower left corner of the matrix U, from bottom to top, from left to right, and eliminates all non-diagonal elements of the matrix. Each step of Givens matrix corresponds to the target matrix to be eliminated, and its four non-zero elements are:

\begin{equation}
\begin{cases}
g_{jj}=\frac{u_{jk}^*}{\sqrt{|u_{ik}|^2+|u_{jk}|^2}}\\
g_{ji}=\frac{u_{ik}^*}{\sqrt{|u_{ik}|^2+|u_{jk}|^2}}\\
g_{ij}=\frac{-u_{ik}}{\sqrt{|u_{ik}|^2+|u_{jk}|^2}}\\
g_{ii}=\frac{u_{jk}}{\sqrt{|u_{ik}|^2+|u_{jk}|^2}}
\end{cases}
\end{equation}

Givens rotation needs to be further decomposed. One idea is to use Gray code \cite{Barenco1995Elementary} to decompose it into several $C^{n-1}NOT$ gates and $C^{n-1}U$ gates. Another method using GCB encoding does not need $C^{n-1}NOT$ gate, but only needs one $C^{n-1}U$ gate to implement Givens rotation acting on n quantum bits. As shown in Figure 5(a), the core idea of this method is to rearrange the target matrix, so that the bit codes corresponding to the adjacent basis vectors of the matrix only differ by one quantum bit, which corresponds to the target quantum bit of $C^{n-1}U$ gate. Therefore, there is no need to flip through $C^{n-1}NOT$ gate. We set $c_i$ as the serial number of the matrix, and $i$ corresponds to the decimal representation of $|i\rangle$, and the encoding method is:

\begin{equation}
c_i=i {\rm xor} (i/2)
\end{equation}

Finally, it should be pointed out that not every control bit in the Givens matrix corresponding to $C^{n-1}U$ gate is necessary \cite{vartiainen2004efficient}, as shown in Figure 5(c), we only need to keep those control bits that correspond to the new action after deletion, and the matrix will not have non-zero elements in the rows and columns that have been cleared.

\begin{figure}[!htbp]
    \centering
    \includegraphics[width=0.5\textwidth]{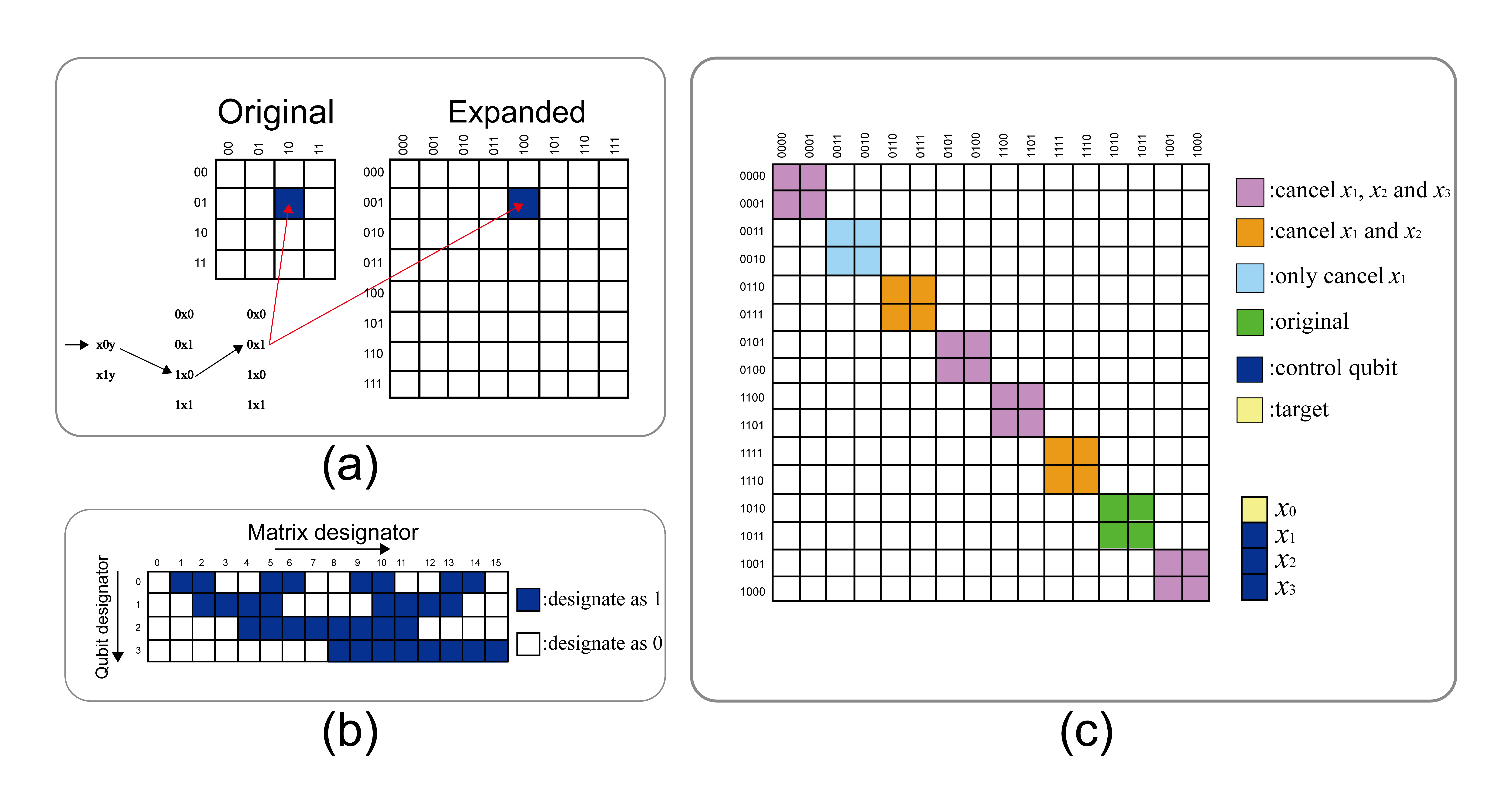}
    \caption{The implementation process of QRD algorithm. Figure (a): The process of rewriting a matrix corresponding to a unitary operation acting on several quantum bits as a matrix form corresponding to a unitary operation acting on the whole system. In QRD implementation process, here we list the case of extending two quantum bits to four quantum bit system, which is a three-channel structure; Figure (b): The GCB encoding method in four quantum bit system, where only one quantum bit ringing between two basis vectors has different color; Figure (c): The process of canceling control bits, which will cause non-zero elements to appear in different positions, and it is necessary to judge whether a control bit can be canceled according to different positions.}
\end{figure}

The purpose of Barenco configuration introduced before is to implement the decomposition of multi-bit controlled $U$ gate. This configuration recursively decomposes a $C^nU$ gate into a $C^{n-1}U$ gate, a $CU$ gate and a $C^{n-1}NOT$ gate from top to bottom. The end point of recursion is a $CNOT$ gate or a $CU$ gate. $CNOT$ gate belongs to universal quantum gate set, while $CU$ gate needs to be further decomposed into $CNOT$ gate and single quantum bit gate. The single quantum bit gates in Barenco configuration are decomposed into three quantum gates rotating around Bloch sphere coordinate axis by ZY decomposition or XY decomposition. These methods can be connected with QRD algorithm results.

\subsection{QSD algorithm}

QSD algorithm, also known as quantum Shannon decomposition algorithm \cite{shende2005synthesis}, is a method that uses linear algebra and optimization theory to decompose any unitary operation into a combination of single bit rotation gates and controlled not gates. QSD is based on a construction called quantum multiplexer, which can decompose any unitary operation into basic quantum gates by decomposing the unitary operation into quantum multiplexers and implementing different quantum multiplexers. The process of decomposition by stages is shown in Figure 6. QSD algorithm can provide a powerful and flexible tool for quantum computation field, and promote the implementation and testing of quantum algorithms on actual hardware. Further improving the efficiency of quantum compilation also requires QSD algorithm or QRD algorithm as the basis.

\begin{figure}[!htbp]
    \centering
    \includegraphics[width=0.5\textwidth]{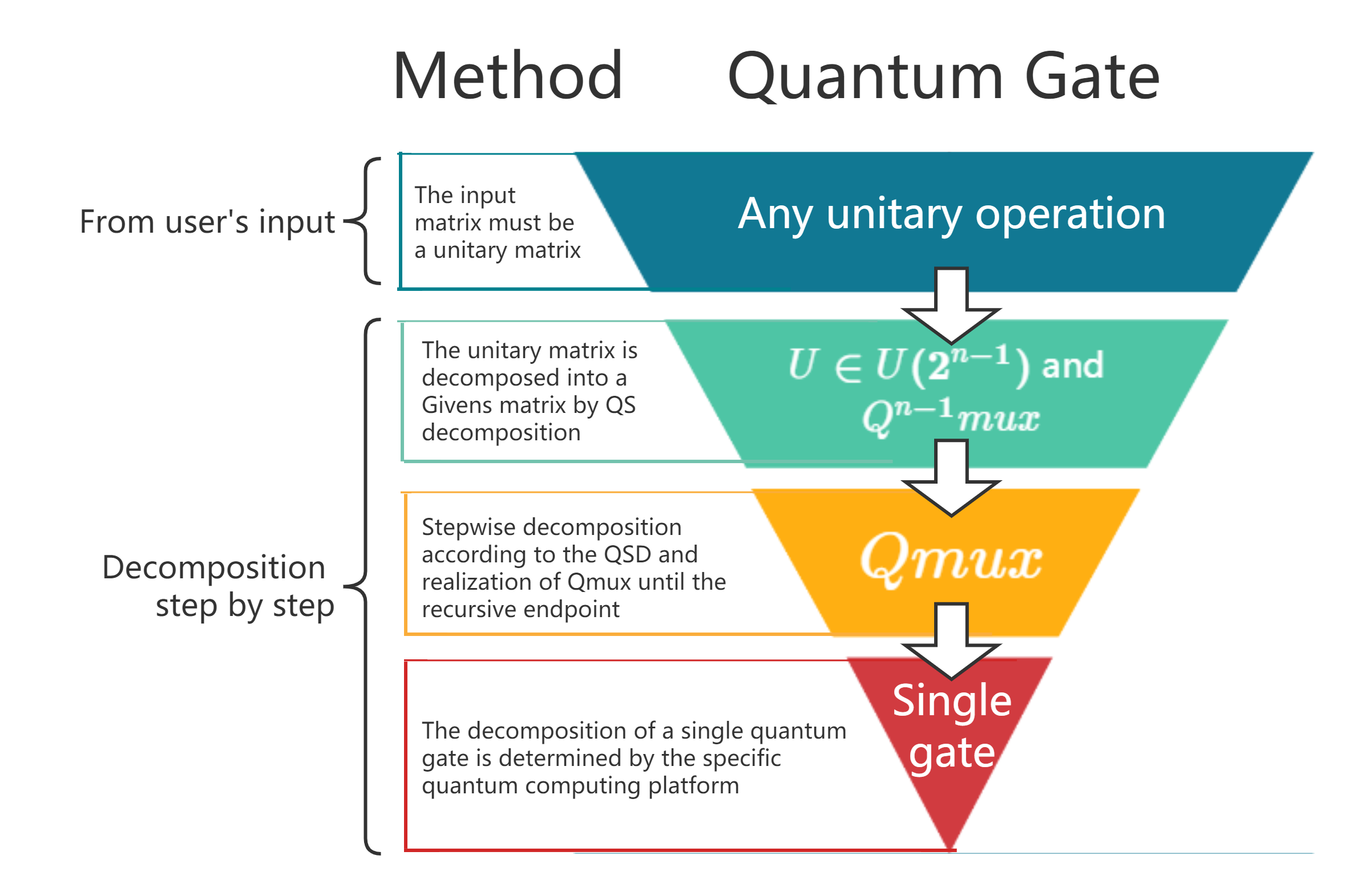}
    \caption{The process of implementing a unitary operation decomposition using QSD algorithm. The whole process is a process of decomposition by stages.}
\end{figure}

Quantum multiplexer is a kind of quantum logic circuit, which can select the state of one or more target quantum bits according to the state of one or more control quantum bits. Qmux can be regarded as the quantum generalization of classical multiplexer, which can implement various quantum operations, such as quantum selection, quantum exchange, quantum permutation, etc. Qmux is an effective quantum logic circuit design tool, which can help to implement universal quantum computation and quantum information processing. A quantum multiplexer divides the quantum bits in the quantum circuit into three categories: action quantum bits, control quantum bits and irrelevant quantum bits. Quantum multiplexer as a quantum gate means that for different combinations of basis vectors of control quantum bits, the quantum gate acting on action quantum bits is different. Irrelevant quantum bits do not affect the whole process, and the result of action will be directly tensor product to the unitary space of irrelevant quantum bits.

\begin{figure}[!htbp]
    \centering
    \includegraphics[width=0.5\textwidth]{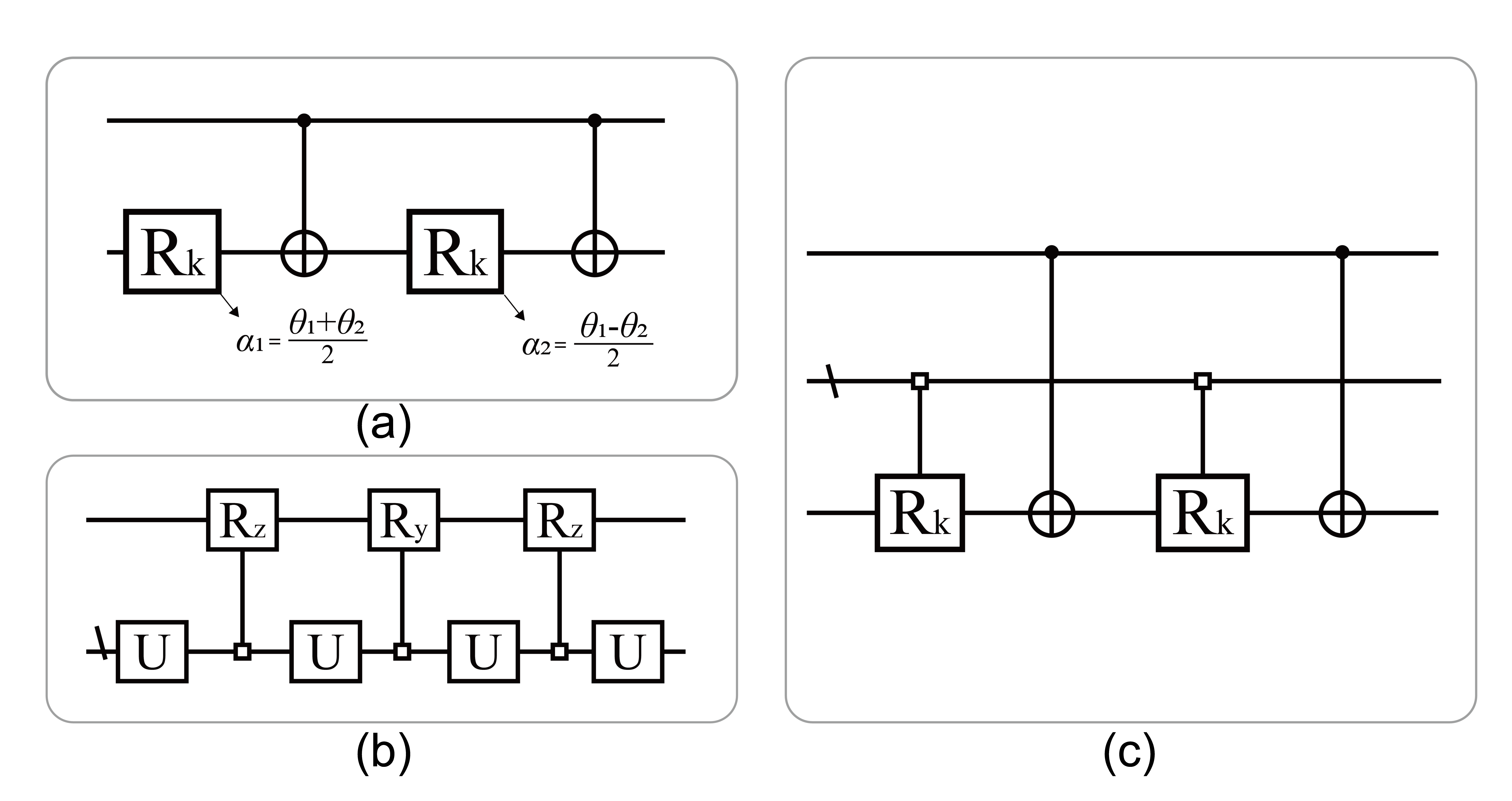}
    \caption{The implementation process of QSD algorithm. Figure (a): The implementation of a rotation form quantum multiplexer with only one control bit, the rotation axes of the front and back rotation gates are the same, and the rotation angles $\alpha_1$ and $\alpha_2$ satisfy $\alpha_1=\frac{\theta_1+\theta_2}{2}$ and $\alpha_2=\frac{\theta_1-\theta_2}{2}$; Figure (b): The principle of quantum Shannon decomposition, which is also a recursive implementation structure. The decomposed quantum multiplexer with multiple control bits can be implemented according to Figure (c). The $2^{n-1}$ order matrix obtained can be recursively processed by quantum Shannon decomposition until the end point of recursion. The end point of recursion can be the optimal form of quantum circuit with any $n$ quantum bits. Here we take the optimal construction with two quantum bits; Figure (c): The implementation of a rotation form quantum multiplexer with multiple control bits.}
\end{figure}

According to CSD algorithm, a unitary matrix can be decomposed into the following form:

\begin{equation}
U=
\begin{pmatrix}
A_1&0\\
0&B_1\\
\end{pmatrix}
\begin{pmatrix}
C&-S\\
S&C\\
\end{pmatrix}
\begin{pmatrix}
A_2&0\\
0&B_2\\
\end{pmatrix}
\end{equation}

For the matrices on both sides, they can be represented as two quantum multiplexers with only one control bit. For the middle matrix, $C,S$ represent sine matrix and cosine matrix respectively, and their forms are:

\begin{equation}
\begin{cases}
C={\rm diag}(\cos\theta_0,...,\cos\theta_{n-1})\\
S={\rm diag}(\sin\theta_0,...,\sin\theta_{n-1})\\
\end{cases}
\end{equation}

This matrix can be represented as a rotation quantum multiplexer around $Y$ axis with $n-1$ control bits, which can be recursively decomposed according to the above implementation scheme. The matrices on both sides can be further decomposed into the following form:

\begin{equation}
\begin{pmatrix}
A&0\\
0&B\\
\end{pmatrix}=
\begin{pmatrix}
V&0\\
0&V\\
\end{pmatrix}
\begin{pmatrix}
D&0\\
0&D^{\dagger}\\
\end{pmatrix}
\begin{pmatrix}
W&0\\
0&W\\
\end{pmatrix}
\end{equation}

In the above equation, the matrices on the left and right sides represent quantum gates on $n-1$ quantum bits, and the middle one corresponds to a rotation quantum multiplexer around $Z$ axis. Summarizing the above equations, we can get the final result of quantum Shannon decomposition, which is a typical recursive structure, and the end point of recursion can be arbitrarily chosen as double quantum bit gate or triple quantum bit gate, etc. By giving the optimal double quantum bit gate or triple quantum bit gate, we can get an optimization. The bottom of QSD algorithm is to find the optimal configuration of quantum gates on a small number of quantum bits \cite{shende2003minimal}, for double quantum bit gate, that is, to find the optimal circuit topology on quantum circuit with two quantum bits, and use the least quantum gates to implement all unitary operations on double quantum bits. We show the implementation of the three most important quantum circuits of the QSD algorithm in Figure 7, where Figure 7(a)(b)(c) correspond to the realization of three different levels of quantum multiplexers.

\subsection{Adaptation method for neutral atom quantum computing platform}

In the neutral atom system, the universal quantum gate sets \cite{xue2006universal} are $C(\theta,\phi)$ gate, $CZ$ gate and $CCZ$ gate, so the algorithms mentioned before cannot be directly applied to this system. We need to find an efficient method to replace the quantum gates in the quantum circuit generated by the matrix decomposition algorithm. The key problem is the conversion relationship between single quantum bit gates.

\begin{figure}[!htbp]
    \centering
    \includegraphics[width=0.5\textwidth]{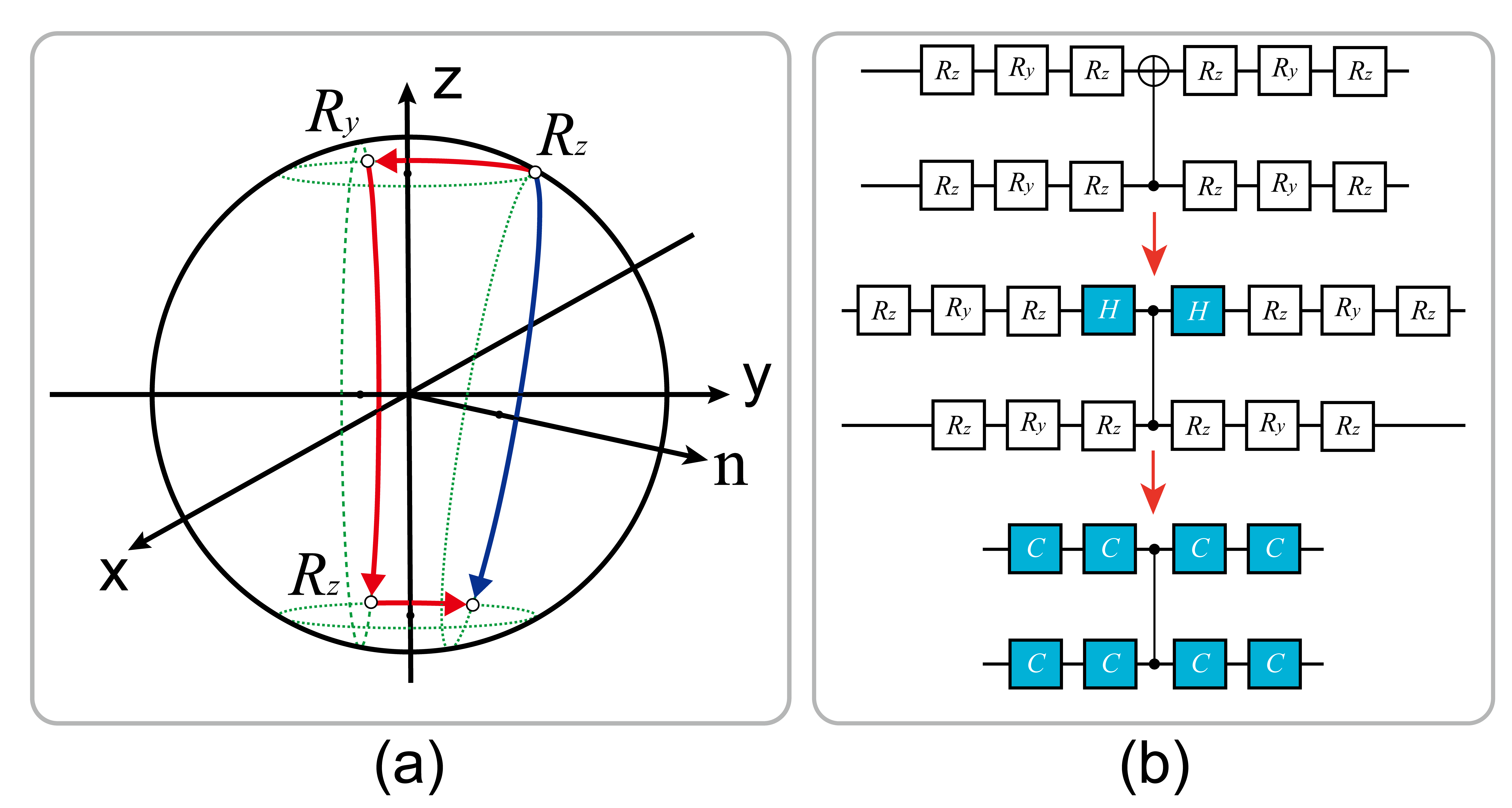}
    \caption{The adaptation of quantum computation in neutral atom system. Figure (a): The decomposition of single quantum bit gate, which is to find an equivalent rotation on the Bloch sphere. Through three consecutive rotations, the three red arrows in the figure indicate the effect of three consecutive rotations, and the blue arrow indicates the effect of equivalent rotation. The calculation process is to find the $\vec n$ and $\theta$ corresponding to the blue rotation; Figure (b): The quantum circuit obtained by matrix decomposition compilation is replaced by two steps with fewer single quantum bit gates and can be executed on the neutral atom quantum computation platform. We demonstrate the process of doing this replacement on a quantum circuit with two quantum bits in the figure. The new quantum bit gates replaced are marked with blue boxes.}
\end{figure}

We can decompose any single quantum bit operation into the following form \cite{martinez2016compiling}:

\begin{equation}
U=C(\theta_2,\phi_2)C(\theta_1,\phi_1)
\end{equation}

To calculate the four parameters in the formula, we introduce the concept of quaternion, which corresponds both axis rotation quantum gate and single quantum bit gate to a quaternion representation of rotating a certain angle around a certain axis. The calculation process is equivalent to using quaternion to correspond arbitrary axis rotation $R_n(\theta)$ and a rotation formed by a continuous combination of three rotations.

Quaternion is a mathematical structure that can be used to rotate objects in three-dimensional space and perform other types of calculations. It consists of a real part and three imaginary parts, usually expressed as $q = w + xi + yj + zk$, where $w,x,y,z$ are real numbers, $i,j,k$ are imaginary units (satisfying $i^2 = j^2 = k^2 = ijk = -1$).

Generally, we will split a quaternion into the following form:

\begin{equation}
q=[\cos(\frac{\theta}{2}),\sin(\frac{\theta}{2})(u,v,w)]
\end{equation}

This formula can be used to represent a rotation with an angle of $\theta$ around a unit vector $(u,v,w)$. Here we assume that $u^2+v^2+w^2=1$. Therefore, we can use quaternion to represent the rotation around the coordinate axis, that is, $R_x(\theta),R_y(\theta),R_z(\theta)$ as follows:

\begin{equation}
\begin{cases}
q_x=[\cos(\frac{\theta}{2}),\sin(\frac{\theta}{2})(1,0,0)]\\
q_y=[\cos(\frac{\theta}{2}),\sin(\frac{\theta}{2})(0,1,0)]\\
q_z=[\cos(\frac{\theta}{2}),\sin(\frac{\theta}{2})(0,0,1)]
\end{cases}
\end{equation}

In quaternion, continuous multiplication represents continuous action of rotation, and the result of multiplication corresponds to the total rotation effect of each sub-rotation. In our decomposition, we first decompose a unitary matrix operation into three rotations around $Z,Y$ axes. If we want to find out the total rotation corresponding to this rotation is how much angle around which axis, we can multiply the four quaternions corresponding to three rotations in turn, and then decompose the result into the form of quaternion above.

ZY decomposition corresponds to the rotation mentioned above as $q_zq_yq_z$, which is the result of multiplying three quaternions in turn. We split the resulting quaternion into the basic form of quaternion $[\cos(\frac{\theta}{2}),\sin(\frac{\theta}{2})(u,v,w)]$, and rewrite the rotation axis $(u,v,w)$ as $(\cos\phi\sin\beta,\sin\phi\sin\beta,\cos\beta)$. We set the parameters of the two $C(\theta,\phi)$ in the decomposition as $\theta_1=\theta_2=\theta$ and $\phi_1=\phi+\frac{\Delta}{2},\phi_2=\phi-\frac{\Delta}{2}$, and substitute the corresponding results into the calculation to get the corresponding results:

\begin{equation}
\begin{cases}
\cos^2\frac{\beta}{2}=\frac{1}{2}(\cos\frac{\alpha}{2}+1)\sin^2\theta\\
\cos\Delta=\frac{\cos^2\frac{\beta}{2}-\cos\frac{\alpha}{2}}{1-\cos^2\frac{\beta}{2}}
\end{cases}
\end{equation}

The above results give the specific values of $\theta_1,\theta_2,\phi_1,\phi_2$, which correspond to the time and phase of the laser applied to the alkali metal atoms in the neutral atom quantum computation system.

Rewriting the controlled gate as suitable for neutral atom quantum computation is relatively simple. We can now naturally derive the conversion relationship between two universal quantum gate sets. Now we only need to replace $CNOT$ in Barenco configuration with $CZ$, and Toffoli gate with $CCZ$.

In the actual algorithm execution process, we implement the replacement of the above quantum gates in two steps. We give an example in Figure 8(b). The operation of replacing $CNOT$ gate and Toffoli gate with $CZ$ gate and $CCZ$ gate is executed in the matrix decomposition algorithm process. After the algorithm execution, we will get a quantum circuit with several single quantum bit rotation gates, Hadamard gates generated by replacement, and $CZ$ gate and $CCZ$ gate. After that, I will execute the second step, which is to replace the single quantum bit operation in the quantum circuit with two $C(\theta,\phi)$ gates by using the quaternion method mentioned before. When executing the replacement of single quantum bit gate, first combine the single quantum bit operations that act continuously on the same quantum bit into a single quantum bit operation $U$, and then execute the replacement to reduce the number of single quantum bit gates in the circuit.


\section{Effect of quantum compilation algorithm}

\subsection{The compilation effect of pyQPanda}

At present, many quantum computing related software provide the interface of quantum compilation algorithm bases on matrix decomposition. We use pyQPanda\cite{dou2022qpanda} developed based on Python, which provides the interface of matrix decomposition to decompose the unitary operation corresponding to the unitary matrix into a series of quantum gates. The interface requires wto parameters, the first one is all the quantum bits used, and the second one is the unitary matrix to be decomposed. The algorithm is to decompose a $2^n$ order unitary matrix into one more than $r=\frac{N(N-1)}{2}$ single quantum logic gates with a small amount of control, where $N=2^n$. The unitary matrix we use in the experiment is randomly genrated in the $U(2^n)$ set. Since the most expensive quantum gate in the neutral atom quantum computing platform is the entanglement quantum gate, the complexity of the circuit is based on the number of entanglement gates as the statistical indicator.

\begin{figure*}[!htbp]
    \centering
    \includegraphics[width=0.8\textwidth]{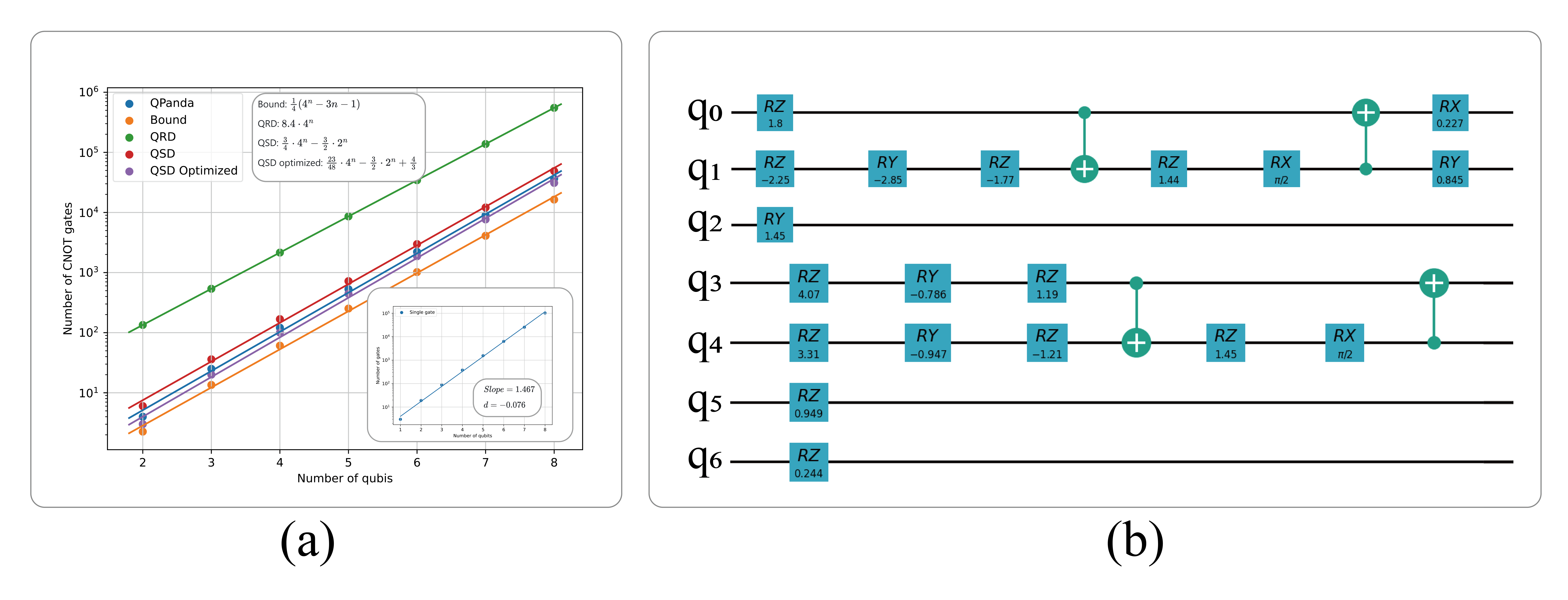}
    \caption{Quantum compilation based on matrix decomposition. Figure (a): The number of $CNOT$ gates in quantum circuits generated by different quantum compilation methods based on matrix decomposition increases with the number of quantum bits. The number of single quantum bit gates generated by QPanda and the number of quantum bits are also shown in the figure; Figure (b): The local structure of the quantum circuit obtained by QPanda's quantum compilation, which is a case with seven quantum bits.}
\end{figure*}

The efficiency of different algorithms for decomposition varies, but under the premise of exact decomposition, the lower bound of the number of $CNOT$ gates contained in the quantum circuit implemented by any quantum compilation algorithm is $\frac{1}{4}(4^n-3n-1)$\cite{shende2003minimal}. As shown in Figure 9(a), the quantum circuit generated by QPanda's quantum compilation is between the general QSD algorithm and the QSD algorithm with low-level optimization, and there is still a gap from reaching the theoretical lower bound. As shown in Figure 9(b), due to the different universal quantum gate sets, the quantum circuit generated by QPanda cannot be directly executed on the neutral atom quantum computing system. Table 1 of reference 29 gives the decomposition efficiency of different algorithms\cite{krol2022efficient}.

\subsection{Adaptation effect}

In order to achieve the previous goal, we transform pyQPanda generated quantum circuit into a quantum circuit with fewer single quantum bit gates and can be executed on neutral atom quantum computing platform. We take pyQPanda generated quantum circuit as an example for adaptation operation. First, we generate multiple random $2^n$ dimensional unitary matrices, corresponding to a quantum circuit with $n$ quantum bits. We use pyQPanda to decompose the unitary matrix into basic quantum gates. pyQPanda provides three decomposition methods: QSD, CSD and QRD. According to the previous test results, we know that QSD method obtains a quantum circuit with smaller complexity, so we choose QSD as our decomposition method.

\begin{figure*}[!htbp]
    \centering
    \includegraphics[width=0.8\textwidth]{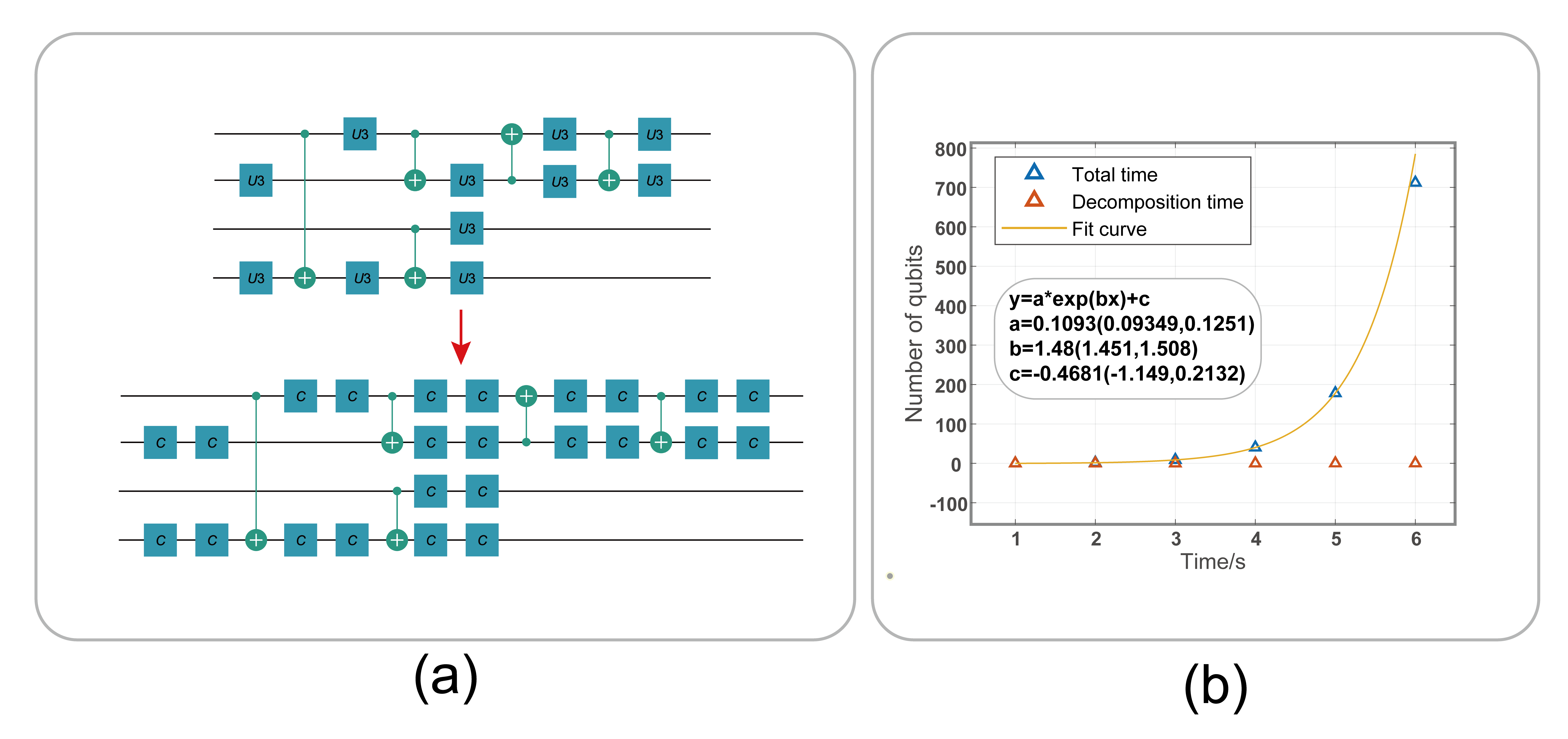}
    \caption{The effect of quantum compilation algorithm for neutral atom quantum computing platform based on quaternion method. Figure (a): Through the adaptation algorithm, we can get a quantum circuit suitable for neutral atom quantum computing platform. Here we give the local replacement result of the circuit; Figure (b): The time complexity analysis of the algorithm. The algorithm has an exponential time complexity with the number of quantum bits, which is the exponential wall problem that cannot be avoided in quantum compilation.}
\end{figure*}

pyQPanda generated quantum circuit is in OPENQASM 2.0 format. We transform it into SEQUENCE format suitable for our processing, execute the corresponding adaptation algorithm, and replace the $CNOT$ gate and single quantum bit gate in it. We have uploaded the Python-based compilation algorithm for neutral atom quantum computing on Github, the website is: https://github.com/Moke2001/QuantumCompilerForUltrocoldAtoms.

We show the adaptation effect in Figure 10, where (a) shows that the initial circuit's $U3$ gate cannot be executed directly. In addition, as shown in Figure 10(b), we analyze the time complexity of our algorithm. The time complexity of our algorithm is different from that of the decomposition algorithm by a large constant, but they are both exponential, which is the exponential wall problem that cannot be avoided in quantum compilation. We point out here: if we can integrate our adaptation into pyQPanda, then the time consumption of the adaptation algorithm will be greatly reduced, and even close to the time consumption of the decomposition algorithm.


\section{Conclusion}

\subsection{Work review}

This paper proposes a quantum compilation algorithm based on matrix decomposition for neutral atom quantum computing platform, as an important component of universal quantum computer. The algorithm can efficiently decompose any unitary operation into a sequence of quantum gates suitable for neutral atom platform, and ensure that the generated quantum circuit can be directly executed on the platform.

Based on the method proposed in this paper, we can implement any unitary operation on the “Hanyuan” series quantum computing platform, and test the potential of neutral atom quantum computing system as a universal quantum computer. In addition, this method can also provide valuable reference for the physical implementation of quantum gates, including laser calibration, etc. At the same time, for the implementation of more complex quantum circuits, we can reduce their complexity based on quantum compilation method, so that the circuit can be executed on quantum computer to get meaningful results.

\subsection{Future prospects}

At present, the lower bound of the number of CNOT gates required for accurate decomposition of a unitary operation is known \cite{shende2003minimal}, but there is no known quantum compilation algorithm that can achieve this lower bound. Finding such an optimal algorithm is still an open problem, which is worth further research. Meanwhile, based on the compilation scheme proposed in this paper, we will explore more quantum algorithm applications and calibration methods on neutral atom quantum computing platform, in order to improve the performance and reliability of the platform.


\bibliography{aipsamp}

\end{document}